\documentclass[11pt]{article} 
\usepackage{csvsimple}
\usepackage[top=2cm, bottom=2cm, left=2cm, right=2cm]{geometry}
\usepackage{natbib}
\usepackage{titlesec}
\usepackage{lipsum}
\usepackage{tipa}
\usepackage{graphicx}
\graphicspath{ {Images/} }
\usepackage{float}
\usepackage{multicol}
\usepackage{float}
\usepackage{multirow}
\usepackage{caption}
\usepackage{fancyhdr}
\usepackage{authblk}
\usepackage{makecell}
\usepackage{changepage}
\usepackage{amsmath} 
\usepackage{subfig}
\usepackage{setspace}
\usepackage{amsfonts}

\titleformat{\section}
{\normalfont\Large\bfseries}
{\thesection\hskip 9pt\textpipe\hskip 9pt}
{0pt}
{}
\begin{document}
\vspace{-2.5cm}
\title{\textbf{The Australian Identification, Nowcasting and Tracking Algorithm (A.I.N.T.)}}

\author{\textsc{\large{Jordan Brook, Hamish McGowan, Matt Mason, Joshua Soderholm and Alain Protat}}}
\affil{In Association with Guy Carpenter LLC.}
\date{12 May 2019}
\maketitle

\vspace{-1.cm}

\begin{center}
\noindent\rule{6cm}{1pt}
\end{center}
\vspace{-0.2cm}
\section{Methodology}

\subsection{Identification}

The storm cell identification method employed in this project starts by identifying spatially contiguous regions above a global field threshold in a vertical maximum reflectivity field ($F$). A minimum area threshold is then applied to each contiguous region to eliminate any anomalously small, high reflectivity regions that arise from non-meteorological origins. The algorithm then checks whether the cell is present at an elevated altitude (4km) to ensure it is not a result of ground clutter at low elevations.\\

This method alone is observed to perform adequately for isolated thunderstorms; however, additional considerations must be made to seperate storm cells embedded within areas of high reflectivity, such as mesoscale convective systems or squall lines \citep{Dixon93}. As a result, each initial cell is then subjected to a secondary subcell identification method which aims to separate any distinct storm cells that are present within the original minimum field threshold boundary. This is done by applying an automatically adjusting (hysteresis) threshold ($H$) that is defined relative to the maximum field value within each cell ($M$). \\

The hysteresis threshold in this project is defined by multiples of a set distance ($D$) from the maximum cell field value; namely, $H(n) = M - nD$, where $\{n \in \mathbb{Z} : n > 0\}$. The algorithm then iterates through ascending values of $n$. At each step the following conditions are validated:

\begin{enumerate}
    \item The hysteresis threshold is at least $D$ greater than the minimum field threshold, or equivalently: $ \left(H(n) - F\right)\geq D$ 
    \item No new subcells larger than the minimum area threshold are created when the original cell area is checked for contiguous regions with field values above $H(n)$ 
\end{enumerate}

\begin{figure}[H]%
    \centering
    \subfloat[\normalsize{Simple cell identification}] {{\includegraphics[width=1\linewidth]{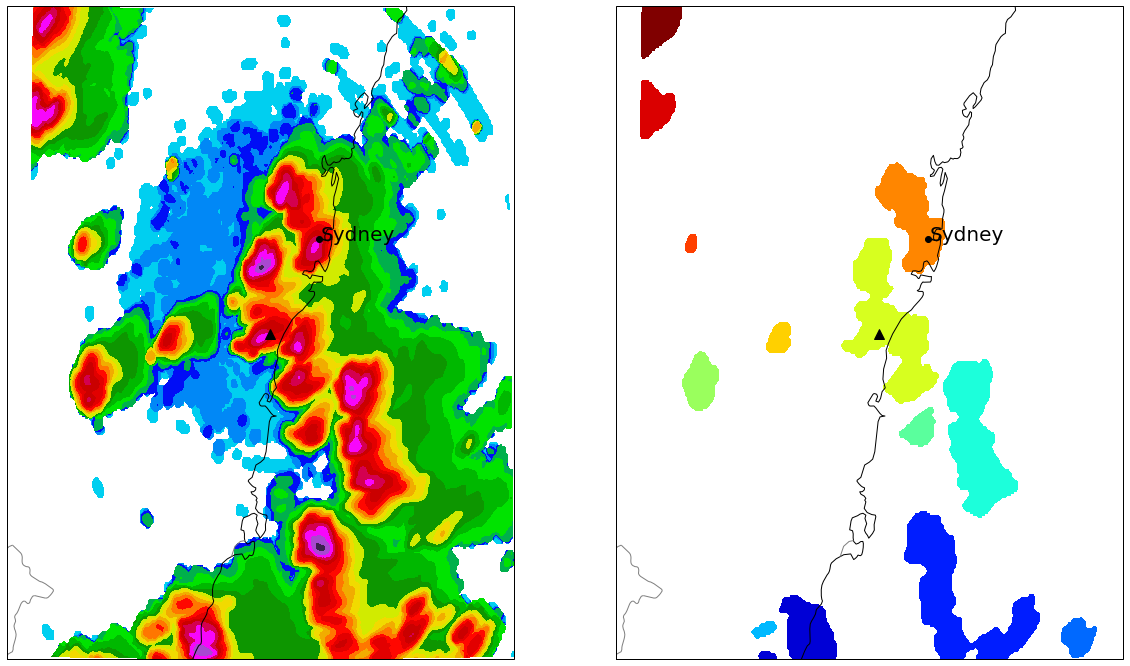} }}%
    \qquad
    \subfloat[\normalsize{Cell identification used in this project}] {{\includegraphics[width=1\linewidth]{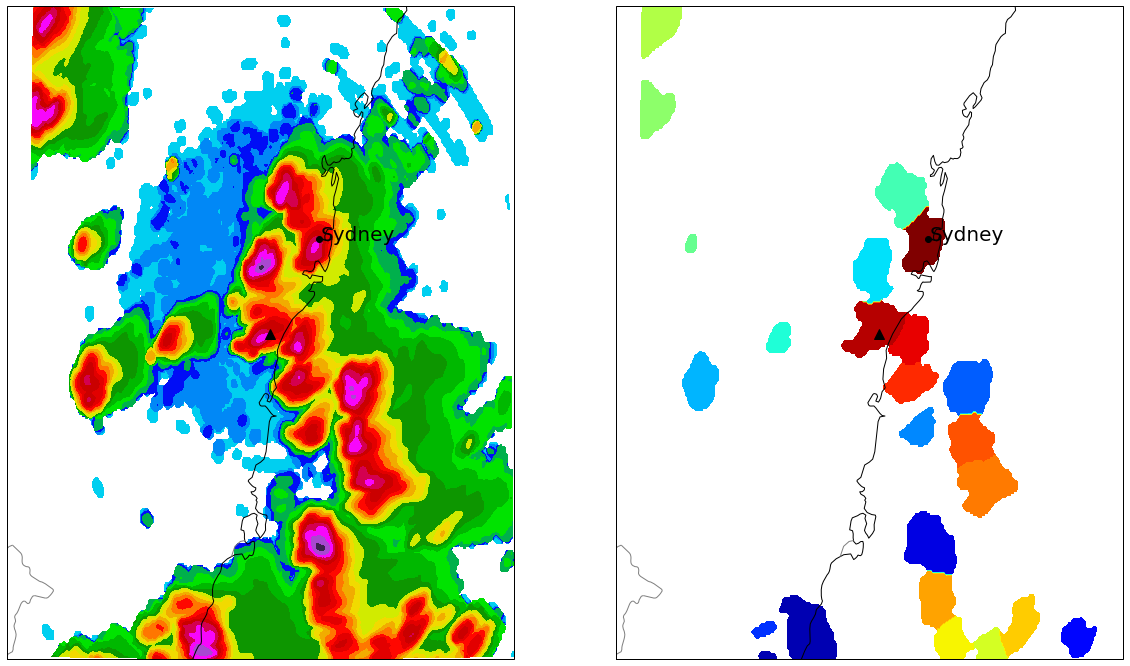} }}%
    \caption{Left: radar reflectivity values for the 20/12/2018 Sydney thunderstorm. Right: A cell identification visualisation where a different color is used for each individual cell.}%
    \label{cells}%
\end{figure}

If condition 1 is broken, the original cell is deemed to contain no additional subcells and is simply outlined by the original minimum field threshold. If condition 2 is broken, new subcells are defined within the original cell region and are outlined by $H(n)$. To retain the same overall field threshold for all cells, subcells must be expanded out to the minimum field threshold boundary. The expansion of subcells is done by assigning any non-labelled grid points inside the original cell boundary to the closest subcell as defined by the geodesic distance. A comparison between simple threshold identification described in the first paragraph and the two-step identification approach used in this project is given in Figure \ref{cells}.

\subsection{Cell Matching and Tracking}

Once the first storm cells in an active thunderstorm period have been identified at time ($t$), they are initialised and given a unique identifying number (UID). If cells are present in the next radar scan at time ($t+1$) the algorithm will attempt to match cells that persist between subsequent scans. The first step in the matching process is to identify a search area at time ($t+1$) in which to search for cells identified at time ($t$). The centre of the search area for a cell at time ($t+1$) may be shifted from its position at time ($t$) as thunderstorms commonly move between radar scans. To quantify this displacement vector, the velocity of the thunderstorm must be estimated at time ($t$). \\

A fast Fourier Transform (FFT) phase correlation technique is used to estimate the movement of cells \citep{Leese71,Johnson98}. In summary, the FFT technique is used to estimate the displacement vector between two similar images that contain a spatial offset relative to one and other. Using FFT phase correlation, a local displacement vector is obtained by comparing the local region around a cell at time ($t$) and ($t+1$) and a global displacement vector is calculated by comparing the whole reflectivity field at time ($t$) and ($t+1$). The last piece of information needed to estimate the optimum displacement vector for each cell is the previous displacement between times ($t-1$) and ($t$). Note that if a new cell is identified at time ($t$), then the previous displacement vector does not exist. The following cases detail which value is returned as the optimum displacement as the three underlying displacement values vary:

\begin{itemize}

    \item \textbf{Case 0:} new cell, local displacement and global displacement disagree\footnote{Agreement between two displacement values is defined as the Euclidean distance between the displacement vectors exceeding a maximum displacement disparity threshold}, returns global displacement
    \item \textbf{Case 1:} new cell, local displacement and global displacement agree, returns local displacement
    \item \textbf{Case 2:} local displacement disagrees with previous displacement and global displacement, returns previous displacement
    \item \textbf{Case 3:} local displacement agrees with global displacement but disagrees with previous displacement, returns local displacement
    \item \textbf{Case 4:} local displacement and previous displacement agree, returns average of both
    \item \textbf{Case 5:} local cell regions empty or at edge of frame, returns global displacement

\end{itemize}

Once the search extent in time ($t+1$) is positioned based on the optimum displacement, all cells within are compared to the original cell at time ($t$). The comparison process involves computing the difference in area ($\Delta A$), position ($ \Delta P$) and bearing ($\Delta B$) between the original cell and the potential match. A disparity metric ($C$) is then defined as follows:

\begin{equation}
\label{cost}
    C = 5\Delta P + 10\Delta A + 20(\Delta B)^{1.5} + \Delta P\times \Delta B
\end{equation}

The numerical constants defined in this disparity metric were refined through trial and error to produce roughly equal disparity for typical changes in area, bearing and position. The change in bearing term was given an added exponential weighting to especially penalise any cell matches that would indicate the cell significantly changed direction. A disparity matrix is created based on Equation (1) between all cells at time ($t$) and time ($t+1$) and the Hungarian optimisation algorithm is used to calculate the optimum set of matches between cells \citep{Dixon93}. \\

Storm mergers occur when initially separate cells merge into one object as defined by the identification algorithm. When this occurs, the new merged cell will retain the unique identifying number (UID) of the original cell that best minimises the disparity metric. Oppositely, when a single storm cell splits into many cells, the new cell which best matches the original will retain the original UID and the other newly split cell/s will be treated as newly initiating storm cells and given new UID's. 

\subsection{Ground Clutter}

The intended use case for the AINT algorithm is on calibrated, quality-controlled radar data such as the level 1b data in the Australian radar archive \citep{Soderholm2019}. These pre-processing efforts aim to limit the existence of any non-meteorological artefacts such as radar noise and ground clutter. However, even with multiple quality control layers, heavy cloud cover moving over elevated topography regions has been observed to cause spurious cell identifications. In an effort to identify where ground clutter is occurring, an added post-processing step has been performed on the AINT thunderstorm archive. The result of the post-processing is to include a clutter flag for each UID that indicates whether the cell is likely to be non-meteorological in origin. \\

\begin{figure}[H]
\centering
\includegraphics[width=0.70\textwidth]{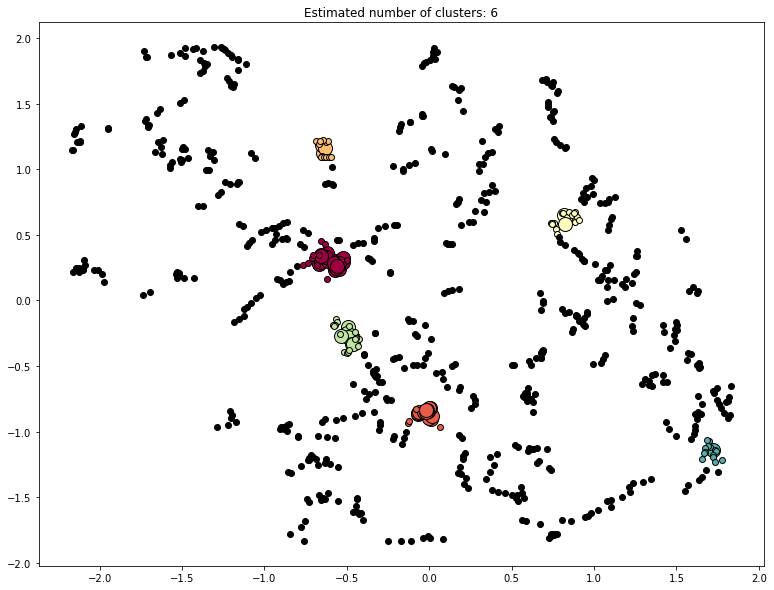}
\caption{An example of a daily accumulation of stationary cells for the Marburg radar. Black points indicate no cluster, and other colours indicate membership to cluster groups.).\label{cluster}}
\centering
\end{figure}

The first step in identifying which cells may be the result of ground clutter is to note any anomalous stationary cells in  otherwise dynamic storm environments. This is done by noting cases where the local displacement (defined in Section 1.2) is close to 0, and the global displacement is significantly different to 0. At the end of each UTC day, the position of all anomalously stationary cells during that period is clustered into spatially defined groups by a density-based spatial clustering of applications with noise (DBSCAN) algorithm \citep{Ester96}. Refer to Figure \ref{cluster} for a visual example of this clustering process. \\

Once a cluster of anomalous stationary cells is identified, a further check is done to identify whether the cluster is positioned in a region where the radar beam is being blocked by elevated topography. The lowest radar beam elevation angle ($\theta$) to hit the surface topography in the vicinity of the radar has been defined as follows: 

\begin{equation}
\theta = arctan\left(\frac{h-h_r}{r}\right) ,
\end{equation}

where $h$ is the height of topography, $h_r$ is the radar altitude and $r$ is the range from the radar. Note that this simplified calculation does not factor in the curvature of the Earth or beam refraction in the atmosphere.  \\

\begin{figure}[H]
\centering
\includegraphics[width=0.70\textwidth]{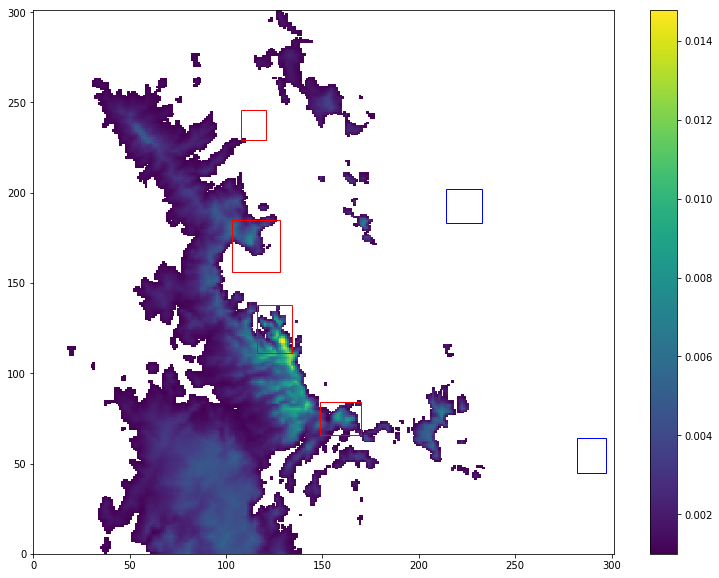}
\caption{Bounding boxes for each cluster in Figure \ref{cluster} plotted with the radar-relative beam angle to the Earth's surface (Equation 2). A 0.001$^{\circ}$ beam angle mask is used to illustrate the high topography region of interest and region boundaries of clusters coincident with high topography are coloured red. \label{topo}}
\centering
\end{figure}

A cluster is defined to be associated with elevated topography if any region within the bounding box surrounding the cluster cells coincides with an elevation angle from topography greater than 0.001$^\circ$. Figure \ref{topo} illustrates an example of this process for the Marburg radar in Queensland, Australia. Finally, once a clutter region has been identified in a 24 hour period, cells that only exist within said region for their lifetimes are flagged as clutter. This ensures cells which either pass through a clutter region or initiate in a clutter region and move away are not labelled as clutter. \\

\begin{figure}[H]%
    \centering
    \subfloat[\normalsize{clutter flag = False}] {{\includegraphics[width=0.75\linewidth]{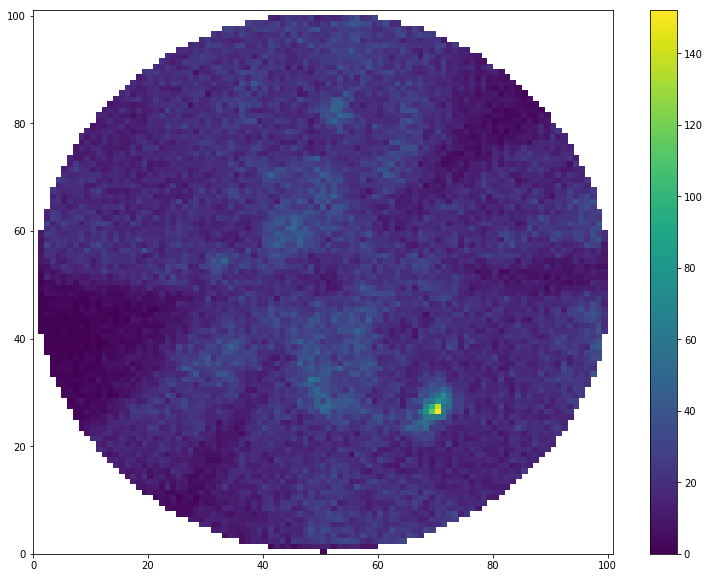} }}%
    \qquad
    \subfloat[\normalsize{clutter flag = True}]{{\includegraphics[width=0.75\linewidth]{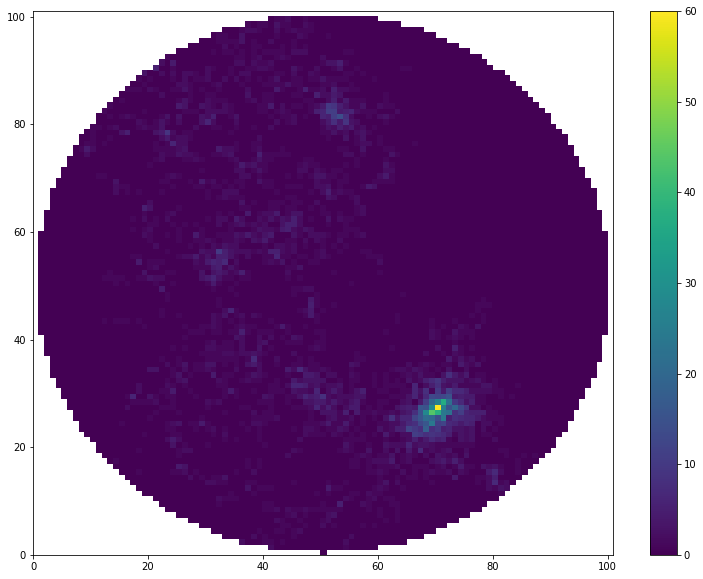} }}%
    \caption{Cumulative counts of thunderstorm area centres that exist for at least 2 radar scans over a 22 year period from 1997 to 2018 for the Marburg radar.  }%
    \label{counts}%
\end{figure}

The ground clutter post-processing algorithm has been developed to perform conservatively and weight the importance of sensitivity (true positive rate) over specificity (true negative rate). This means that cells are only flagged as clutter if they have passed a rigorous series of tests and the confidence in this classification is high. As a result, not all ground clutter is removed by excluding all cells where the clutter flag = True (refer to Figure \ref{counts}). Instead, the clutter flag should be used to assist data users in noting which radars experience significant amounts of clutter and where that clutter occurs. The authors propose that any additional time spent increasing the accuracy of this post-processing step should instead be spent improving the quality controls of the underlying gridded data. These efforts would make the clutter-removal efforts discussed here redundant.

\subsection{MESH Correction}

Radar reflectivity ($Z$) is an exponential property that is proportional to the sixth power of hydrometeor diameter. The exponential nature of reflectivity dictates that large hydrometeors can result in very high $Z$ values. As a more convenient, human-readable measure of reflectivity, this is often transformed to a logarithmic scale measured in dBZ. This transform is given below:

\begin{equation}
    dBZ = 10\cdot log_{10}(Z)
\end{equation}

Interpolation of reflectivity values is required when mapping radar data from spherical coordinates to Cartesian grids. Historically, this interpolation has been done in dBZ units, rather than the more natural $Z$ units \citep{Lakshmanan12}. However, \citet{Warren19} were able to show that interpolating in $Z$ is preferable for severe convection as it more accurately resolves high reflectivity cores. Seeing as this archive is focused on severe convection,  it was deemed appropriate that the underlying reflectivity data here is interpolated in $Z$ coordinates. Whilst the reflectively data in this archive is likely to be much more accurate due to this departure from common methodology, the interpolation method will likely alter other empirically derived radar products. \\

Products such as MESH, POSH and VIL have all been empirically derived using data interpolated in dBZ units. This introduces a bias into the products as data interpolated in dBZ underestimates the true intensity and extent of high reflectively cores. Therefore, when calculations such as MESH are made with data interpolated in $Z$, values will be overestimated. Research is being undertaken to correct this bias by re-calibrating the MESH formula for reflectivity data interpolated in $Z$. In the meantime, due to the importance of MESH values in this archive, a simple linear transform is presented here to convert from biased MESH values to corrected MESH values. This is given below:

\begin{equation}
\label{transform}
MESH_{cor} = 0.88\times MESH_{bias} - 2.9
\end{equation}

\begin{figure}[H]
\centering
\includegraphics[width=0.70\textwidth]{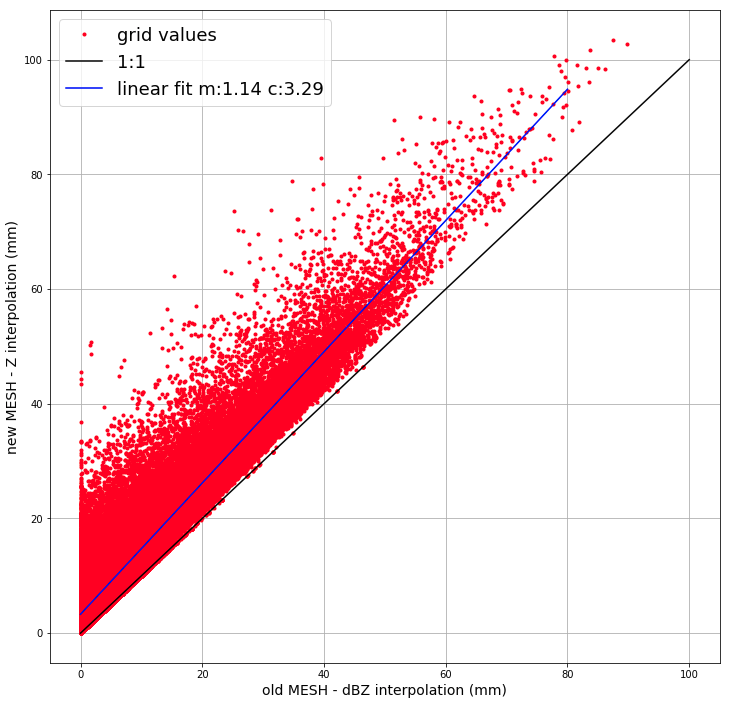}
\caption{Scatter plot showing MESH values calculated using dBZ interpolation vs Z interpolation. The line of best fit represents a linear transform between these two methods. \label{mesh}}
\centering
\end{figure}

The linear transform given above was calculated using all volumes from 00:00UTC - 23:59UTC on large storm days from three spatially disparate radars. These cases are as follows: Melbourne radar 19/12/2017, Marburg radar 16/11/2008 and Wollongong radar 20/12/2018. Figure \ref{mesh} shows a comparison between the two interpolation methods ($Z$ \& dBZ) for all pixels where MESH values exist for all radar volumes on these days. The inverse of the line of best fit between the two interpolation methods from this figure was used to calculate the transform in Equation (\ref{transform}).

\subsection{Archive Parameters}
\setstretch{0.5}
\begin{multicols}{2}

\begin{itemize}
\item \textbf{Field Threshold:} 35 dBZ, field threshold used for cell identification. Detected cells are connected pixels above this threshold.

\item \textbf{Hysteresis Level:} 10 dBZ, multiples of this value (denoted `$D$' in Section 1.1) form the hysteresis threshold used to identify subcells within larger regions above the field threshold.

\item \textbf{Centroid Percentile:} 90 \%, any points with a reflectivity percentile above this percentile threshold are used to calculate the field weighted centroid of each cell.

\item \textbf{Minimum Cell Size:} 100 km$^2$, the minimum size threshold in square kilometers for a cell to be identified.

\item \textbf{Minimum Subcell Size:} 100 km$^2$, same as above except used for defining if a subcell is large enough to be considered a separate cell.

\item \textbf{Search Margin:} 4000 m, the radius of the search box around the predicted object center to look for other cells to match. 

\item \textbf{Flow Margin:} 10000 m, the margin size around the cell extent on which to perform FFT phase correlation to calculate local displacement.

\item \textbf{Maximum Disparity:} 100, maximum allowable disparity value between a possible cell match as defined by Equation (\ref{cost}).

\item \textbf{Maximum Flow Magnitude:} 50 ms$^{-1}$, maximum allowable global displacement magnitude

\item \textbf{Displacement Agreement Threshold:} 10 ms$^{-1}$, maximum magnitude of difference for two displacements to be considered in agreement.

\item \textbf{Cell Check Altitude} : 4000 m, cell identification takes place at this altitude to check that cells exist at this height, before they are expanded back out to the 35 dBZ threshold of the vertical maximum reflectivity field. 

\end{itemize}
\end{multicols}

\setstretch{1}

\subsection{Major Limitations}

\begin{itemize}
    \item \textbf{Ground Clutter:} Radar scans at very low elevations can experence beam blockage from objects such as terrain, buildings, trees etc. The post-processing outlined in Section 1.3 attempts to identify which cell identifications are a result of ground clutter but some spurious cell identifications will not have been flagged by this process.
    \item \textbf{Range Sensitivity:} The effects of beam spreading and attenuation can effect the sensitivity of some radars with range. The recommended usage of data is out to a 150km radius, although this may be less for some older C band radars with wide beamwidths.  
    \item \textbf{Beam Blockage:} Some radars are positioned in areas where a number of scan elevations are permanently blocked by ground clutter. These radars are unlikely to pick up on small, shallow thunderstorms behind this blockage shadow. This is unlikely to effect hail producing storms as they often extend above beam blocked elevations. 
    \item \textbf{MESH in Tropics:} Hail size observations used to formulate MESH were taken in a temperate climate \citep{Witt98}. Caution must be exercised when using MESH values to estimate hail sizes in mid-latitude or tropical regions as the empirical relationship between radar reflectivity and hail size may not hold under these thermodynamic conditions.
\end{itemize}

\section{Output Information}

\subsection{Single Statistics}

Individual storm statistics are calculated for each thunderstorm at each radar scan throughout the entire Australian archive. The following variables are given in daily CSV files for each UTC day that thunderstorms are identified. Daily csv files are named as follows: \textbf{ID\_yyyymmdd\_HHMMSS.storm.csv}\\

\setstretch{0.5}
\begin{multicols}{2}

\begin{enumerate}
\item \textbf{Scan} (integer): Number radar scan between 00:00UTC and 23:59UTC for each day
\item \textbf{UID} (integer): Unique identifier for each cell
\item \textbf{Time} (yyyy-mm-dd hh:mm:ss): Time of the radar scan in UTC time
\item \textbf{Lon} (deg): Longitude of the reflectivity-weighted cell centroid 
\item \textbf{Lat} (deg): Latitude of the reflectivity-weighted cell centroid
\item \textbf{Latlonbox} (min.\ lat, min.\ lon, max.\ lat, max.\ lon): A tuple containing the cell bounding box in lat/lon coordinates
\item \textbf{Area centre x} (float): x position of the area-based centre of the cell in grid coordinates
\item \textbf{Area centre y} (float): y position of the area-based centre of the cell in grid coordinates
\item \textbf{Weighted centre x} (float): x position of the reflectivity-weighted centroid of the cell in grid coordinates
\item \textbf{Weighted centre y} (float): y position of the reflectivity-weighted centroid  of the cell in grid coordinates
\item \textbf{Gridbox} (min.\ y, min.\ x, max.\ y, max.\ x): Tuple containing the cell bounding box in grid coordinates
\item \textbf{Major axis} (deg): Major axis of an ellipse fitted to the cell area in longitude coordinates
\item \textbf{Minor axis} (deg): Minor axis of an ellipse fitted to the cell area in latitude coordinates
\item \textbf{Orientation} (deg): Angle between the ellipse major axis and the positive x-axis. 
\item \textbf{Area} (km$^2$): Area of the 2D footprint of the cell 
\item \textbf{Volume} (km$^3$): Volume of cell 
\item \textbf{Mass} (kt): Total mass of liquid water within cell
\item \textbf{Max refl} (dBZ): Maximum reflectivity value within cell
\item \textbf{Max refl height} (m): Height of the maximum reflectivity value within cell
\item \textbf{Mean refl} (dBZ): Mean reflectivity within the cell bounding box
\item \textbf{Max height} (km): Maximum height of grid points within cell region above the field threshold
\item \textbf{Max top alt} (m): Maximum cloud top height within the field region
\item \textbf{Max sts alt} (m): Maximum 50 dbZ height within cell
\item \textbf{Max VIL} (kgm$^{-2}$): Maximum vertically integrated liquid value within cell region
\item \textbf{Cell VIL} (kgm$^{-2}$): Cell-based vertically integrated liquid value 
\item \textbf{Max MESH} (mm): Maximum MESH value within cell region
\item \textbf{Corrected MESH max} (mm): Maximum corrected MESH value within cell region
\item \textbf{Max POSH} (\%): Maximum POSH value within cell region

\item \textbf{Clutter} (bool): Clutter flag as defined in Section 1.3.

\end{enumerate}
\end{multicols}

\setstretch{1}

\subsection{Gridded Products}

In addition to single statistics, the following gridded products are also calculated and stored in the archive for each storm cell. Gridded data are stored in daily hierarchical data files named as follows: \textbf{ID\_yyyymmdd\_HHMMSS.storm.h5}
\vspace{0.1cm}

\begin{multicols}{2}
\setstretch{0.5}
\begin{enumerate}

\item \textbf{Refl vol} (dBZ): 3D gridded reflectivity volume bounding the cell
\item \textbf{Max refl grid} (dBZ): 2D vertical maximum reflectivity grid
\item \textbf{Cell mask} (bool): 2D grid the same shape as the cell bounding box that outlines the actual cell shape
\item \textbf{Sts h grid} (m): 2D 50 dBZ height grid 
\item \textbf{Top h grid} (m): 2D cloud top height grid 
\item \textbf{VIL grid} (kgm$^{-2}$): 2D vertically integrated liquid grid 
\item \textbf{MESH grid} (mm): 2D maximum estimated size of hail grid
\item \textbf{Corrected MESH grid} (mm): 2D corrected MESH grid
\item \textbf{POSH grid} (\%): 2D Probability of severe hail grid

\end{enumerate}
\end{multicols}
\setstretch{1}

\noindent
Daily HDF files are organised to easily select data from cells listed in the daily CSV files. This data hierarchy is given below:\\

\hspace{0cm} - \textbf{Level 1:} Scan Time (`hhmmss' UTC)

\hspace{1cm} - \textbf{Level 2:} UID 

\hspace{2cm} - Refl vol

\hspace{2cm} - Max refl grid

\hspace{2cm} - Cell mask

\hspace{2cm} - Sts h grid

\hspace{2cm} - Top h grid

\hspace{2cm} - VIL grid

\hspace{2cm} - MESH grid

\hspace{2cm} - Corrected MESH grid

\hspace{2cm} - POSH grid

Lastly, there are a number of important details about the gridded data stored in the file attributes of all daily HDF files. These attributes include but are not limited to: gridding projection, radar latitude/longtiude, grid shape, cartesian grid extent, radar site name etc. These variables are included in the file attributes of every HDF file as they may be useful for users attempting to recreate the underlying grids that the data in Section 2.2 is created on.

\bibliographystyle{apalike}
\bibliography{mendeley.bib}

\begin{thebibliography}{}

\bibitem[Dixon and Wiener, 1993]{Dixon93}
Dixon, M. and Wiener, G. (1993).
\newblock {TITAN: Thunderstorm Identification, Tracking, Analysis, and
  Nowcasting—A Radar-based Methodology}.
\newblock {\em Journal of Atmospheric and Oceanic Technology}, 10(6):785--797.

\bibitem[Ester et~al., 1996]{Ester96}
Ester, M., Kriegel, H.-P., Sander, J., and Xu, X. (1996).
\newblock {A Density-based Algorithm for Discovering Clusters a Density-based
  Algorithm for Discovering Clusters in Large Spatial Databases with Noise}.
\newblock In {\em Proceedings of the Second International Conference on
  Knowledge Discovery and Data Mining}, KDD'96, pages 226--231. AAAI Press.

\bibitem[Johnson et~al., 1998]{Johnson98}
Johnson, J.~T., MacKeen, P.~L., Witt, A., Mitchell, E. D.~W., Stumpf, G.~J.,
  Eilts, M.~D., and Thomas, K.~W. (1998).
\newblock {The Storm Cell Identification and Tracking Algorithm: An Enhanced
  WSR-88D Algorithm}.
\newblock {\em Weather and Forecasting}, 13(2):263--276.

\bibitem[Lakshmanan, 2012]{Lakshmanan12}
Lakshmanan, V. (2012).
\newblock {Image Processing of Weather Radar Reflectivity Data : Should It be
  Done in Z or dBZ ?}
\newblock {\em E-Journal of Severe Storms Meteorology}, 7(3):1--8.

\bibitem[Leese et~al., 1971]{Leese71}
Leese, J.~A., Novak, C.~S., and Clark, B.~B. (1971).
\newblock {An Automated Technique for Obtaining Cloud Motion from
  Geosynchronous Satellite Data Using Cross Correlation}.
\newblock {\em Journal of Applied Meteorology}, 10(1):118--132.

\bibitem[Soderholm et~al., 2019]{Soderholm2019}
Soderholm, J., Protat, A., and Jakob, C. (2019).
\newblock {Australian Operational Weather Radar Dataset}.
\newblock {\em National Computing Infrastructure}.

\bibitem[Warren and Protat, 2019]{Warren19}
Warren, R.~A. and Protat, A. (2019).
\newblock {Should interpolation of radar reflectivity be performed in Z or
  dBZ?}
\newblock {\em Journal of Atmospheric and Oceanic Technology}, 0(0):null.

\bibitem[Witt et~al., 1998]{Witt98}
Witt, A., Eilts, M.~D., Stumpf, G.~J., Johnson, J.~T., Mitchell, E. D.~W., and
  Thomas, K.~W. (1998).
\newblock {An Enhanced Hail Detection Algorithm for the WSR-88D}.
\newblock {\em Weather and Forecasting}, 13(2):286--303.

\end{thebibliography}

\newpage

\appendix

\end {document}